\documentclass{SciPost}

\hypersetup{
    colorlinks,
    linkcolor={red!50!black},
    citecolor={blue!50!black},
    urlcolor={blue!80!black}
}

\usepackage[bitstream-charter]{mathdesign}
\usepackage{siunitx}
\usepackage[italic]{hepnicenames}
\usepackage{subcaption} 
\usepackage{cleveref}
\usepackage{multirow}
\usepackage{booktabs} 

\DeclareSymbolFont{usualmathcal}{OMS}{cmsy}{m}{n}
\DeclareSymbolFontAlphabet{\mathcal}{usualmathcal}

\fancypagestyle{SPstyle}{
\fancyhf{}
\lhead{\colorbox{scipostblue}{\bf \color{white} ~SciPost Physics Proceedings }}
\rhead{{\bf \color{scipostdeepblue} ~Submission }}

\fancyfoot[C]{\textbf{\thepage}}
}

\begin{document}

\pagestyle{SPstyle}

\begin{center}{\Large \textbf{\color{scipostdeepblue}{
Search for flavour-changing neutral current couplings between the top quark and the Higgs boson in multilepton final states with the ATLAS detector\\
}}}\end{center}

\begin{center}\textbf{
Shayma A. H. Wahdan\textsuperscript{1$\star$}, On behalf of the ATLAS collaboration
}\end{center}

\begin{center}

{\bf 1} University of Wuppertal, Wuppertal, Germany

$\star$ \href{mailto:shayma.wahdan@cern.ch}{\small shayma.wahdan@cern.ch}

\end{center}

\definecolor{palegray}{gray}{0.95}
\begin{center}
\colorbox{palegray}{
  \begin{tabular}{rr}
  \begin{minipage}{0.36\textwidth}
    \includegraphics[width=60mm,height=1.5cm]{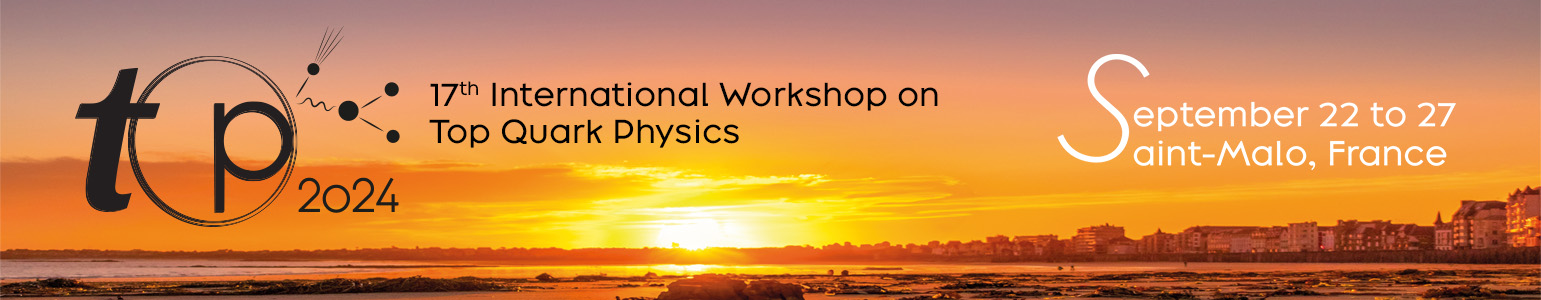}
  \end{minipage}
  &
  \begin{minipage}{0.55\textwidth}
    \begin{center} \hspace{5pt}
    {\it The 17th International Workshop on\\ Top Quark Physics (TOP2024)} \\
    {\it Saint-Malo, France, 22-27 September 2024
    }
    \doi{10.21468/SciPostPhysProc.?}\\
    \end{center}
  \end{minipage}
\end{tabular}
}
\end{center}

\section*{\color{scipostdeepblue}{Abstract}}
\textbf{\boldmath{%
These proceedings present a search for flavour-changing neutral-current (FCNC) interaction involving the top quark, Higgs boson and either the up or the charm quark, using \SI{140}{\mbox{fb\(^{-1}\)}} of \SI{13}{\tera\electronvolt} proton--proton collision data from the ATLAS detector at the Large Hadron Collider. Two channels are considered: the production of top quark-antiquark pair with one top decaying via FCNC, and the associated production of a top quark and Higgs boson. Final states contain either two same-charge leptons, or three leptons of which two have the same charge. Observed (expected) upper limits on the branching rations are determined as \(\mathcal{B}(t\to Hu)<2.8\,(3.0) \times 10^{-4}\) and \(\mathcal{B}(t\to Hc)<3.3\,(3.8) \times 10^{-4}\).
}}

\vspace{\baselineskip}

\noindent\textcolor{white!90!black}{%
\fbox{\parbox{0.975\linewidth}{%
\textcolor{white!40!black}{\begin{tabular}{lr}%
  \begin{minipage}{0.6\textwidth}%
    {\small Copyright attribution to authors. \newline
    This work is a submission to SciPost Phys. Proc. \newline
    License information to appear upon publication. \newline
    Publication information to appear upon publication.}
  \end{minipage} & \begin{minipage}{0.4\textwidth}
    {\small Received Date \newline Accepted Date \newline Published Date}%
  \end{minipage}
\end{tabular}}
}}
}

\section{Introduction}
\label{sec:intro}
Flavour-changing neutral-current (FCNC) interactions are forbidden at tree-level in the Standard Model (SM) and are strongly suppressed at higher orders by the Glashow-Iliopoulos-Maiani (GIM) mechanism~\cite{GIM_mechanism}. The SM predicts extremely small branching ratios for FCNC processes involving the top quark and Higgs boson, with $t \to Hq$ (where $q$ is an up or charm quark) expected to occur at a rate of around $10^{-15}$. However, various beyond the SM theories, such as the two-Higgs-doublet models~\cite{Branco:2hdm2012}, predict significantly enhanced ratios of up to $10^{-3}$. Thus, any observation of these rare interactions at the LHC would indicate the existence of new physics beyond the SM. These proceedings summarize the search of FCNC interactions between the top quark and the Higgs boson in multi-lepton final state~\cite{ATLAS:2024mih}, based on data recorded with the ATLAS detector~\cite{PERF-2007-01} in the years 2015 to 2018 at a centre-of-mass energy of 13 TeV.

\makeatletter
\renewcommand\@makefntext[1]{\noindent#1} 
\footnotetext{Copyright 2024 CERN for the benefit of the ATLAS Collaboration. Reproduction of this article or parts of it is allowed as specified in the CC-BY-4.0 license.}
\makeatother

\section{FCNC search}
\subsection{Signal regions and background estimate}
\begin{figure}[!h]
  \centering
  \begin{subfigure}{0.3\linewidth}
      \includegraphics[width=\linewidth]{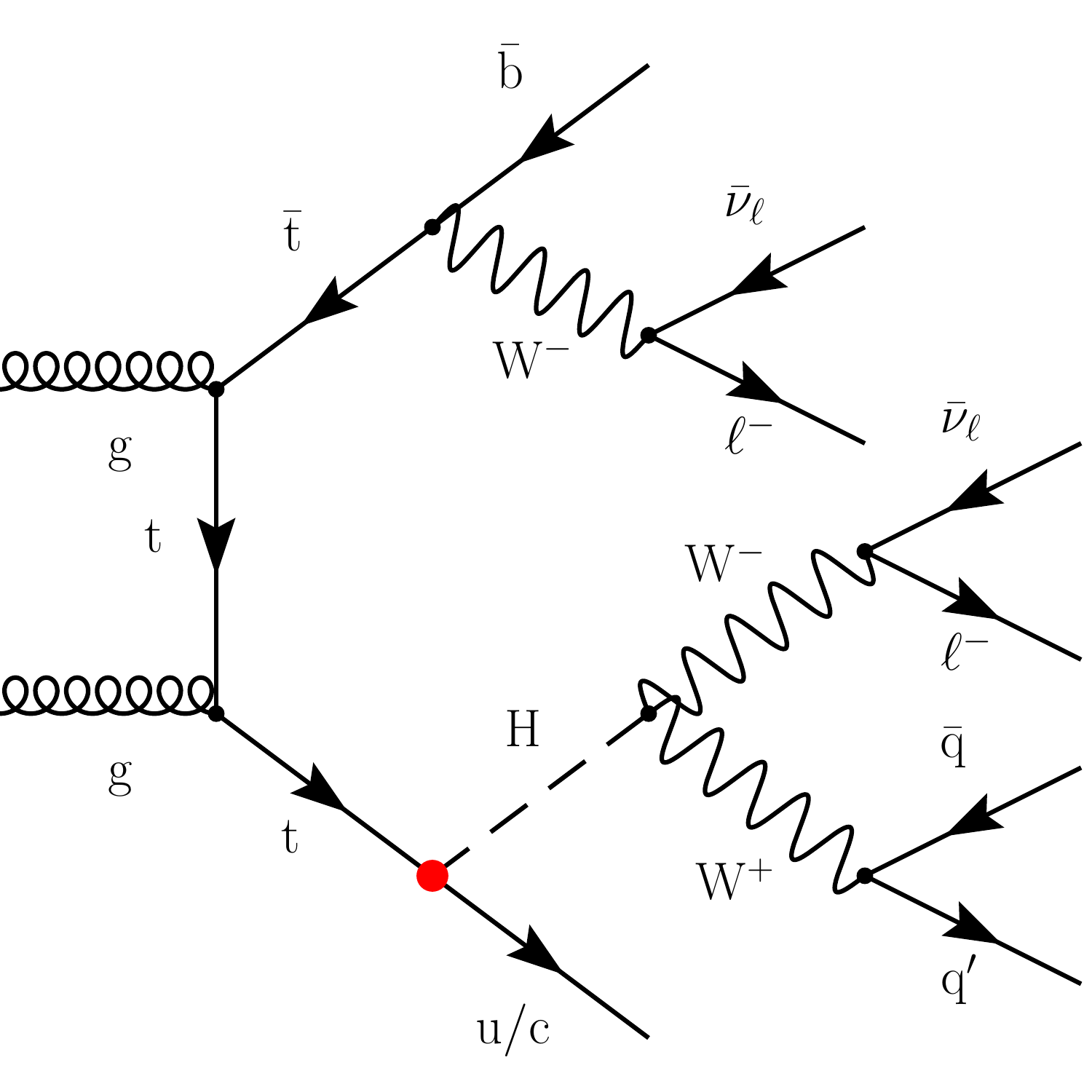}
      \caption{}
  \end{subfigure}
  \hspace{0.15\linewidth}
  \begin{subfigure}{0.3\linewidth}
      \includegraphics[width=\linewidth]{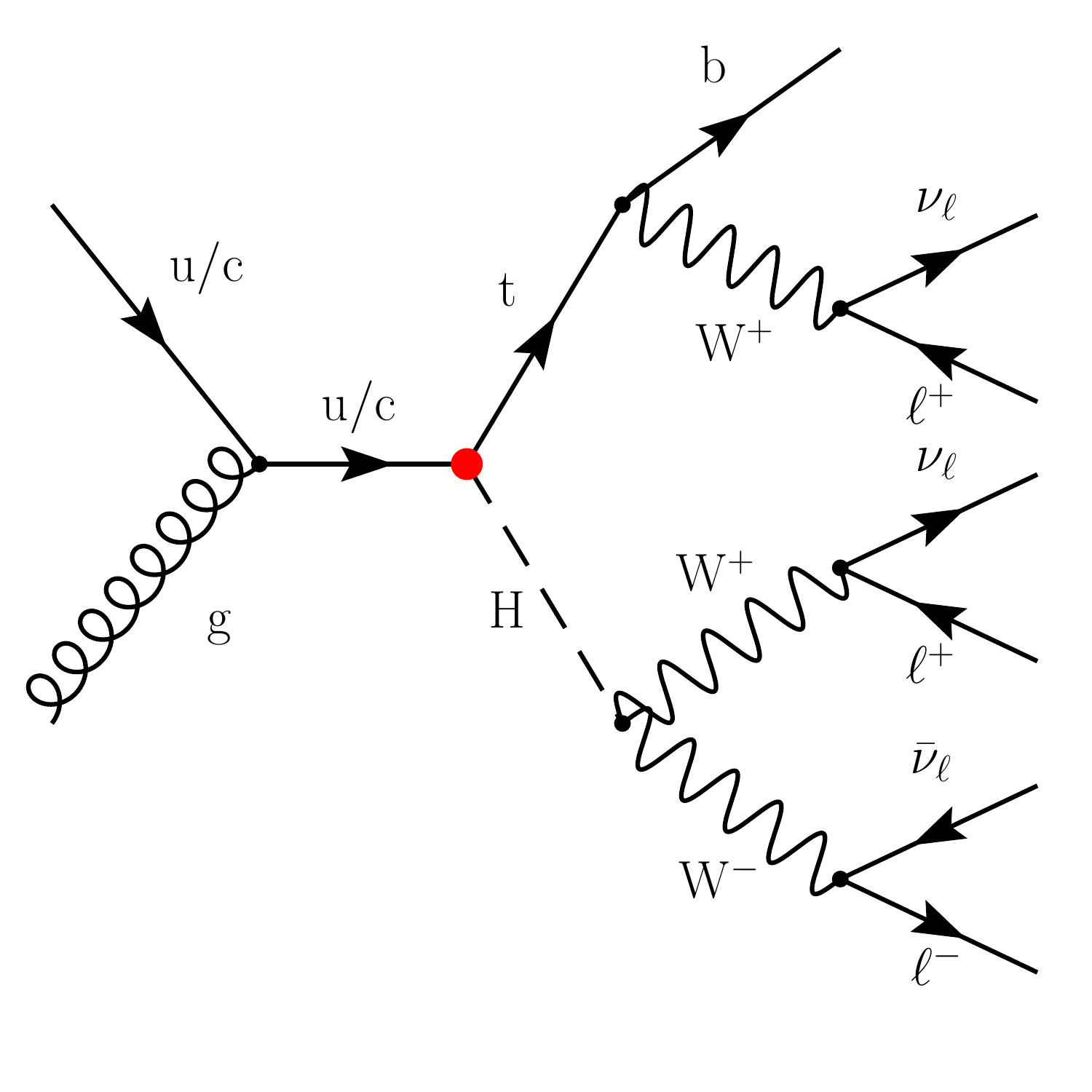}
      \caption{}
  \end{subfigure}
  \caption{
      Example Feynman diagrams of the signal process: 
      (a) \(t\bar{t}(t\to Hq)\) decay resulting in the \(2\Pl SS\) final state, 
      and (b) \(gq\to Ht\) production signal process resulting in the \(3\Pl\) final state.
  }
  \label{fig:feynman_signal}
\end{figure}

Signal regions are defined for FCNC interactions involving the top quark, the Higgs boson, and an up-type quark ($q=u,c$). The considered processes are the $t\bar{t}$ and $Ht$ production.  In $t\bar{t}$ production, one top quark decays via $t \rightarrow Hq$, making it enriched in the $t \rightarrow Hq$ decay signal; this is referred to as the 'Dec' channel. The other $Ht$ production process, on the other hand, contains a larger fraction of $q g \rightarrow tH$ production and is referred to as the 'Prod' channel. The analysis focuses on final states containing either two same-sign leptons ($2\Pl SS, \Pl = e,\mu$) or three leptons ($3\Pl, \Pl = e,\mu$), with the latter requiring exactly two leptons to have the same charge. Example Feynman diagrams are shown in~\Cref{fig:feynman_signal}. 

Events are required to have at least one lepton with $p_{\mathrm{T}} > 28$\,GeV, while additional leptons must have $p_{\mathrm{T}} > 10$\,GeV. Additionally, each event must contain at least one jet, with at least one b-tagged jet to enhance signal purity. Additional selection criteria are applied to define the specific signal regions for the $2\Pl SS$ and $3\Pl$ final states, and to suppress contributions from background processes.

\begin{figure}[!h]
    \centering
    \includegraphics[width=0.9\linewidth]{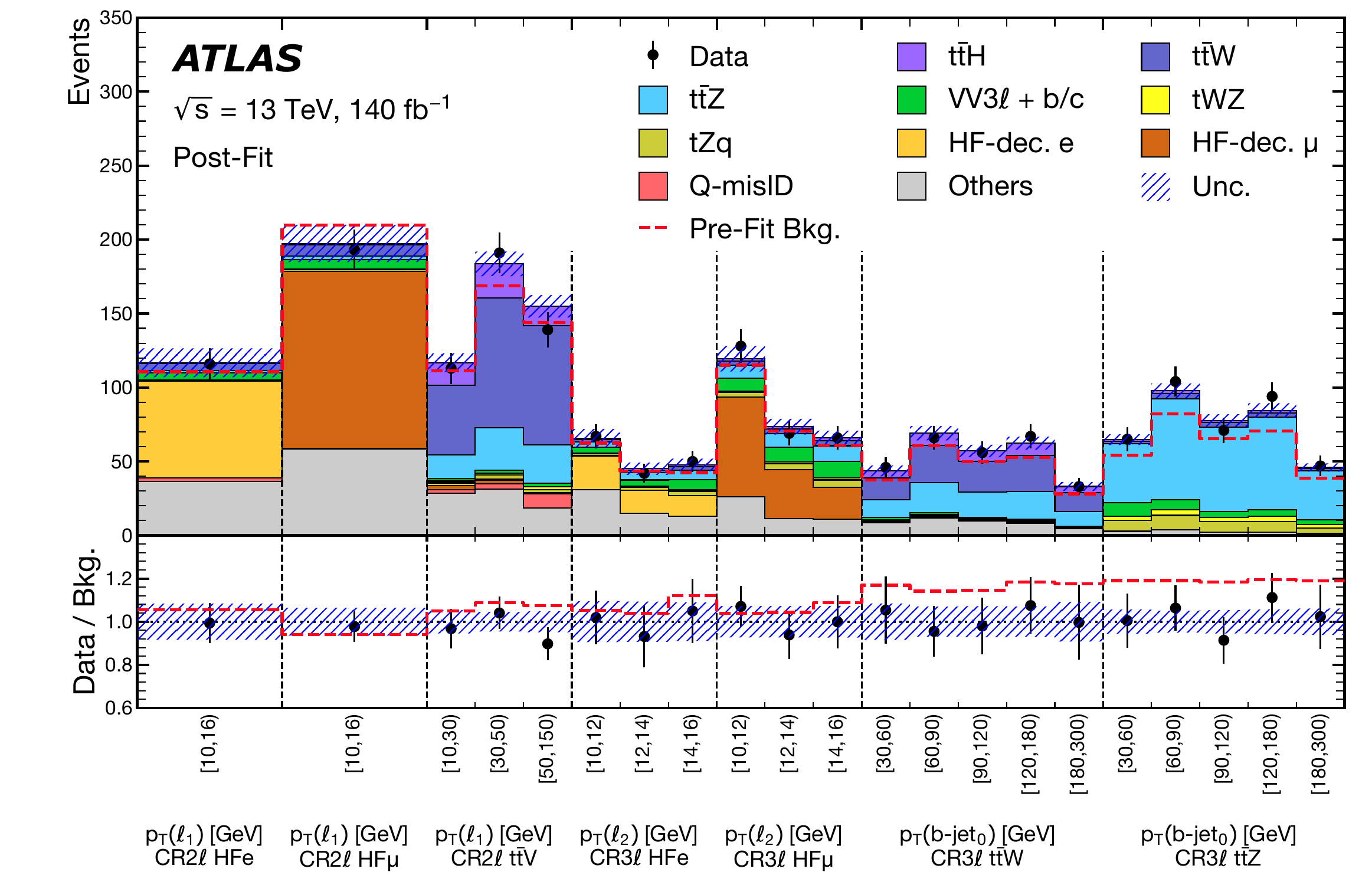}
    \caption{Summary plot of the fitted distributions in all control regions obtained from the signal-plus-background fit to data in the $tHc$ channel.}\label{fig:post_fit_CR_summary}
\end{figure}

Several SM background processes are present in the signal regions, with some processes simulated using Monte Carlo (MC) while others require dedicated treatment. A primary focus is on non-prompt leptons, which predominantly originate from $B$-hadron decays, mainly from $t\bar{t}$ production. The Template Fit Method is employed to estimate the rates of non-prompt leptons background, with normalisation factors determined through simultaneous maximum-likelihood fits in both the two-lepton and three-lepton final states. To ensure the validity of this method, several control regions are defined. Additionally, the $2\Pl SS$ phase space is contaminated with events containing one prompt electrons with misidentified charge; this background, is modeled using a data-driven approach that incorporates events with $SS$ and $OS$ pair of electrons of an invariant mass around the mass of the $Z$-boson. Prompt-lepton backgrounds primarily arise from $t\bar{t} W$ and $t\bar{t} Z$ processes, where the normalisation of these backgrounds is left free-floating in the fit, for which control regions are defined to constrain their contributions. Minor backgrounds and other SM processes are grouped into an "Others" category, as their contributions are similar in the analysis phase space. A summary plot of all $2\Pl$ and $3\Pl$ control regions is shown in~\Cref{fig:post_fit_CR_summary}. Overall, a good agreement between MC and data is observed.

\subsection{Separation of signal and background}

\begin{figure}[!h]
  \centering
  \begin{subfigure}{0.49\textwidth}
      \includegraphics[width=\textwidth]{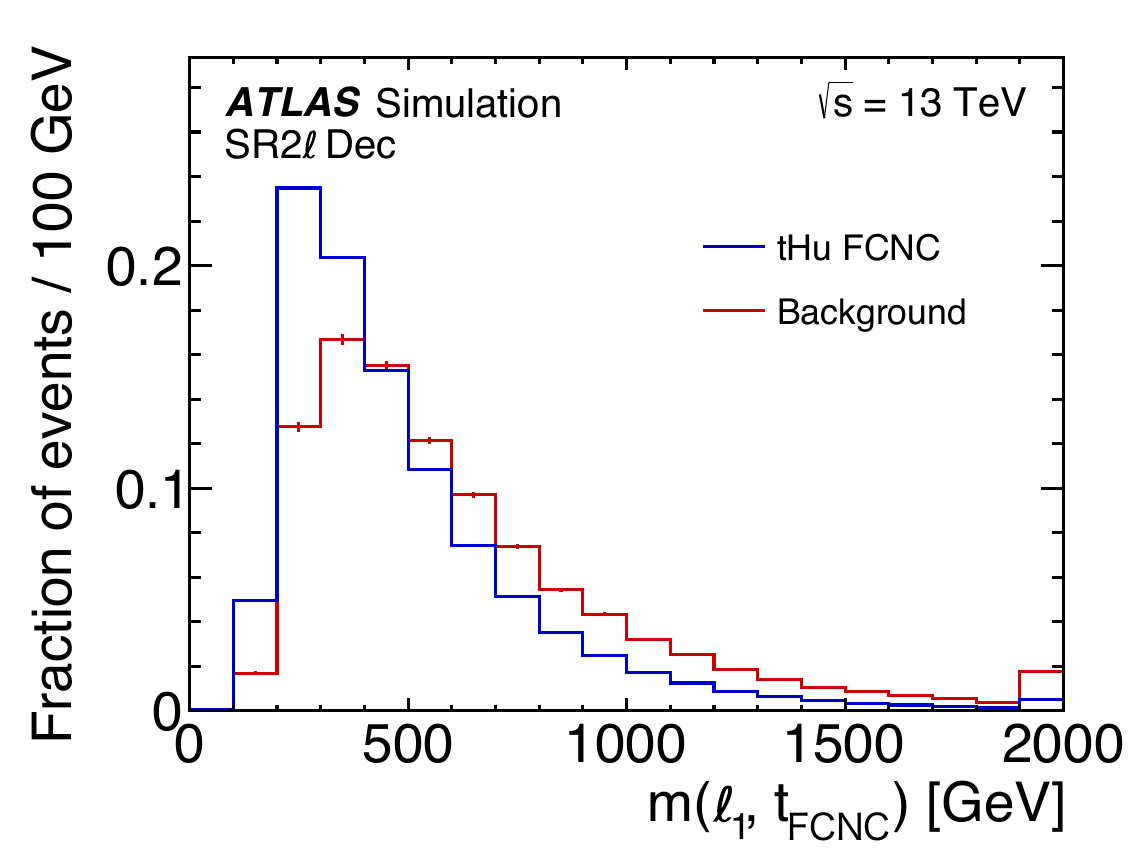}
      \caption{}
  \end{subfigure}
  \hfill
  \begin{subfigure}{0.49\textwidth}
      \includegraphics[width=\textwidth]{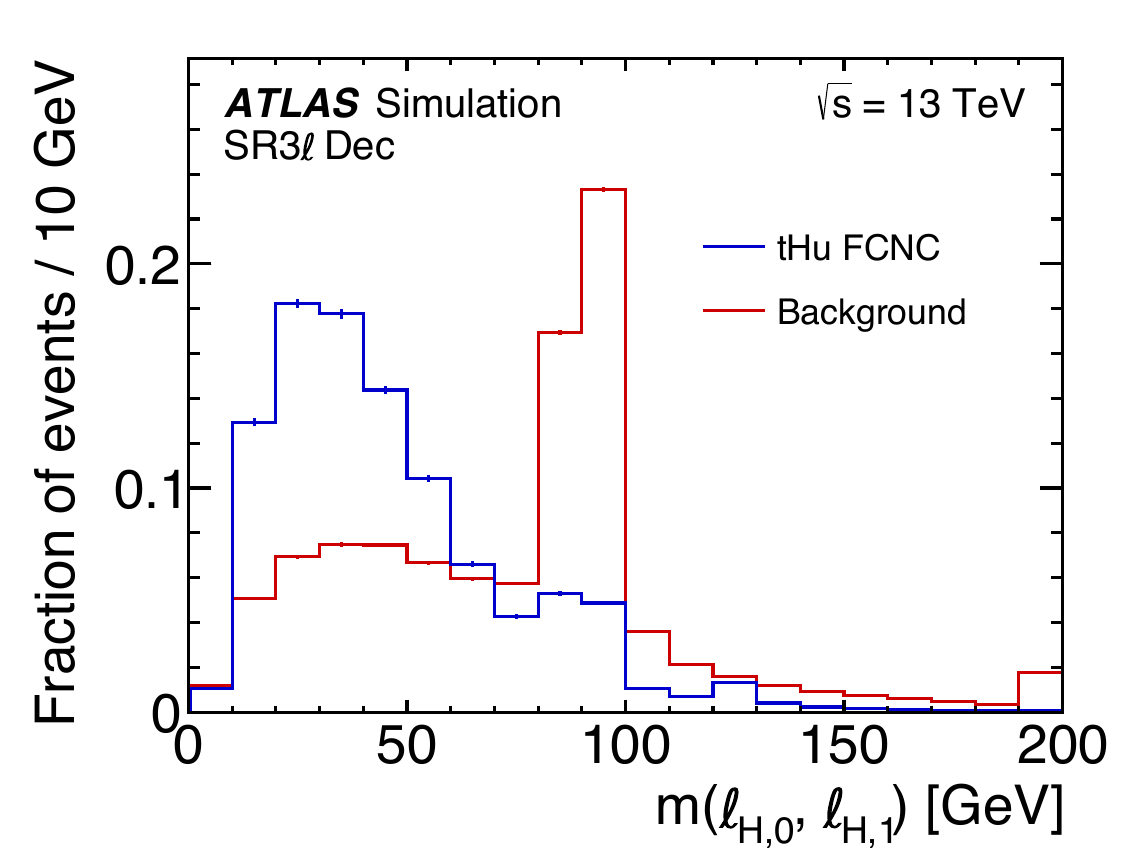}
      \caption{}
  \end{subfigure}
  \caption{
    Probability densities for the \(tHu\) signal process and the sum of all background processes: 
    (a) the mass of the second-highest \(p_{\mathrm{T}}\) lepton (\(\Pl_1\)) and the top quark decaying via \(t \rightarrow Hu\) (\(t_{FCNC}\)) and 
    (b) the mass of the two leptons with the smallest angular separation.
  }
  \label{fig:Reco_example_vars}
\end{figure}

Two custom reconstruction algorithms, Recursive Jigsaw Reconstruction and the Neutrino Independent Combinatorics Estimator, are developed to create variables that enhance the separation between the signal and background. These variables are combined into a single discriminant, $D_{\mathrm{NN}}$ score, using artificial neural networks (NNs). \Cref{fig:Reco_example_vars} shows example of reconstructed variable from each of the reconstruction algorithm, showing clear shape differences between the signal and the combination of all background processes.

\subsection{Results}

\begin{figure}[!h]
  \centering
  \begin{subfigure}{0.4\linewidth}
      \includegraphics[width=\linewidth]{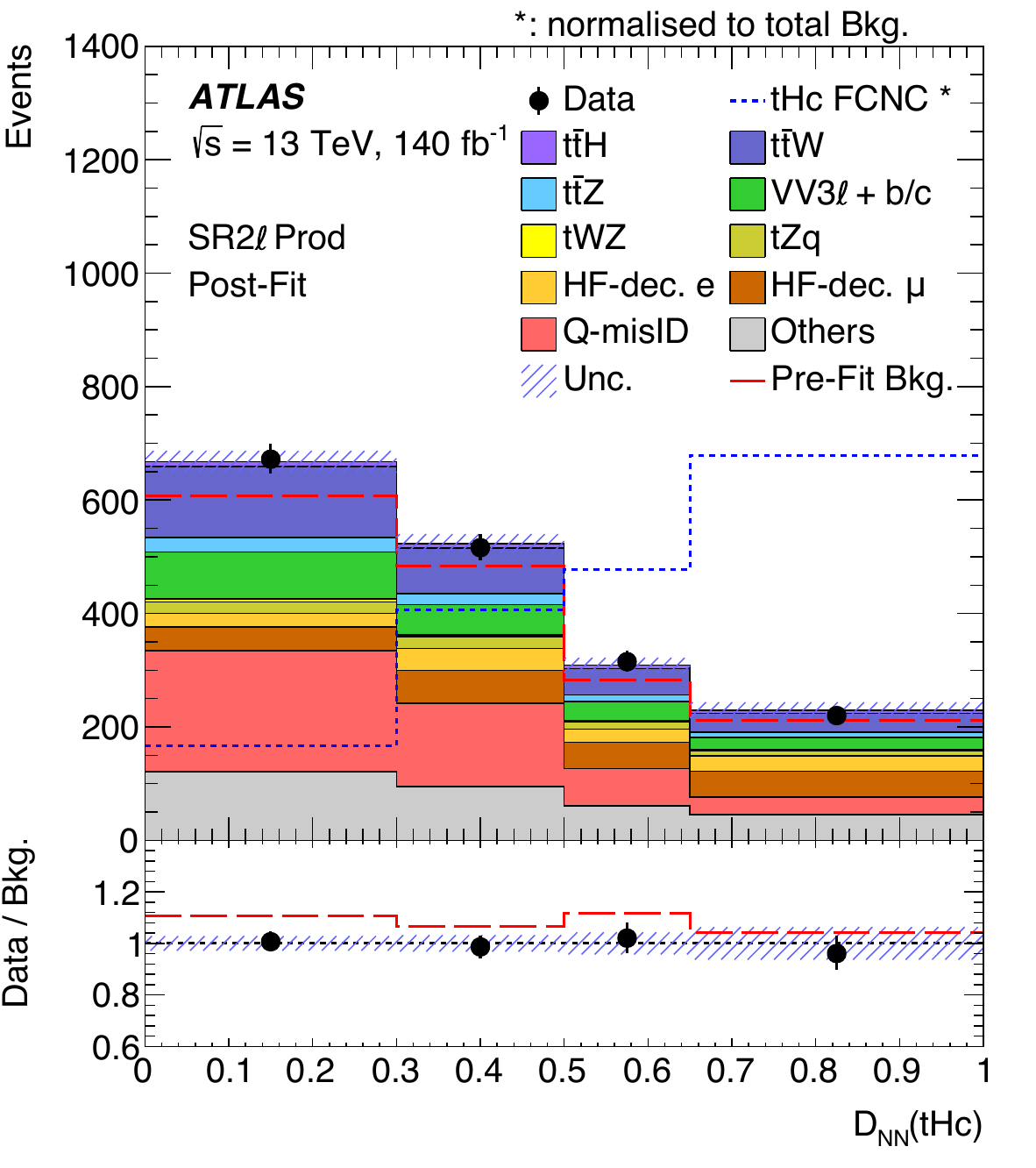}
      \caption{}
  \end{subfigure}
  \caption{
      The \(D_{NN}\) distributions in one of the signal regions obtained from the signal-plus-background fit to data in the \(tHc\) channel. The dotted line represents the distribution of the signal, scaled to the number of background events. The dashed line depicts the sum of all background processes prior to the fit.
  }
  \label{fig:post_fit_NN}
\end{figure}

The NN output distribution is used as input for the profile-likelihood fit to obtain the signal normalisation. Prior to performing a full fit to the data, a background-only fit is performed in regions with low signal sensitivity to ensure a consistent modeling of background processes. The post-fit $D_{NN}$ distribution in one of the signal regions is shown in~\Cref{fig:post_fit_NN}. The best-fit value for the normalization of the $tHu(tHc)$ signal was found to be compatible with zero, showing no significant discrepancies beyond $1\sigma$ for any nuisance parameters. Therefore, upper exclusion limits on the branching ratio \(\mathcal{B}(t\to Hq)\) and the Wilson coefficients of the effective field theory (EFT) dimension-6 operators $C_{u\phi}$ are calculated using the $\mathrm{CL_s}$ method~\cite{Read:2002hq}. \Cref{tab:limits} summarizes the observed and expected upper limits for both the $tHu$ and $tHc$ signals.

\begin{table}[!h]
\centering
\caption{
The observed (expected) upper exclusion limits at \SI{95}{\percent} confidence level~(CL) on the branching ratio \(\mathcal{B}(t\to Hq)\) and the absolute value of the Wilson
coefficient $C_{u\phi}$.
}\label{tab:limits}
    \begin{tabular}{ccc}
        \toprule
        \multirow{2}{*}{Signal} & \multicolumn{2}{c}{Observed (expected) \SI{95}{\percent} CL upper limits }  \\
        & \(\mathcal{B}(t\to Hq)\)            & \(|C_{u\phi}^{qt,tq}|\)  \\
        \midrule
        $tHu$   & \(2.8\,(3.0)\times10^{-4}\) & \(0.71\,(0.73)\) \\
        \midrule
        $tHc$   & \(3.3\,(3.8)\times10^{-4}\) & \(0.76\,(0.82)\) \\
        \bottomrule
    \end{tabular}
\end{table}

\begin{figure}[!h]
  \centering
  \begin{subfigure}{0.49\textwidth}
      \includegraphics[width=\textwidth]{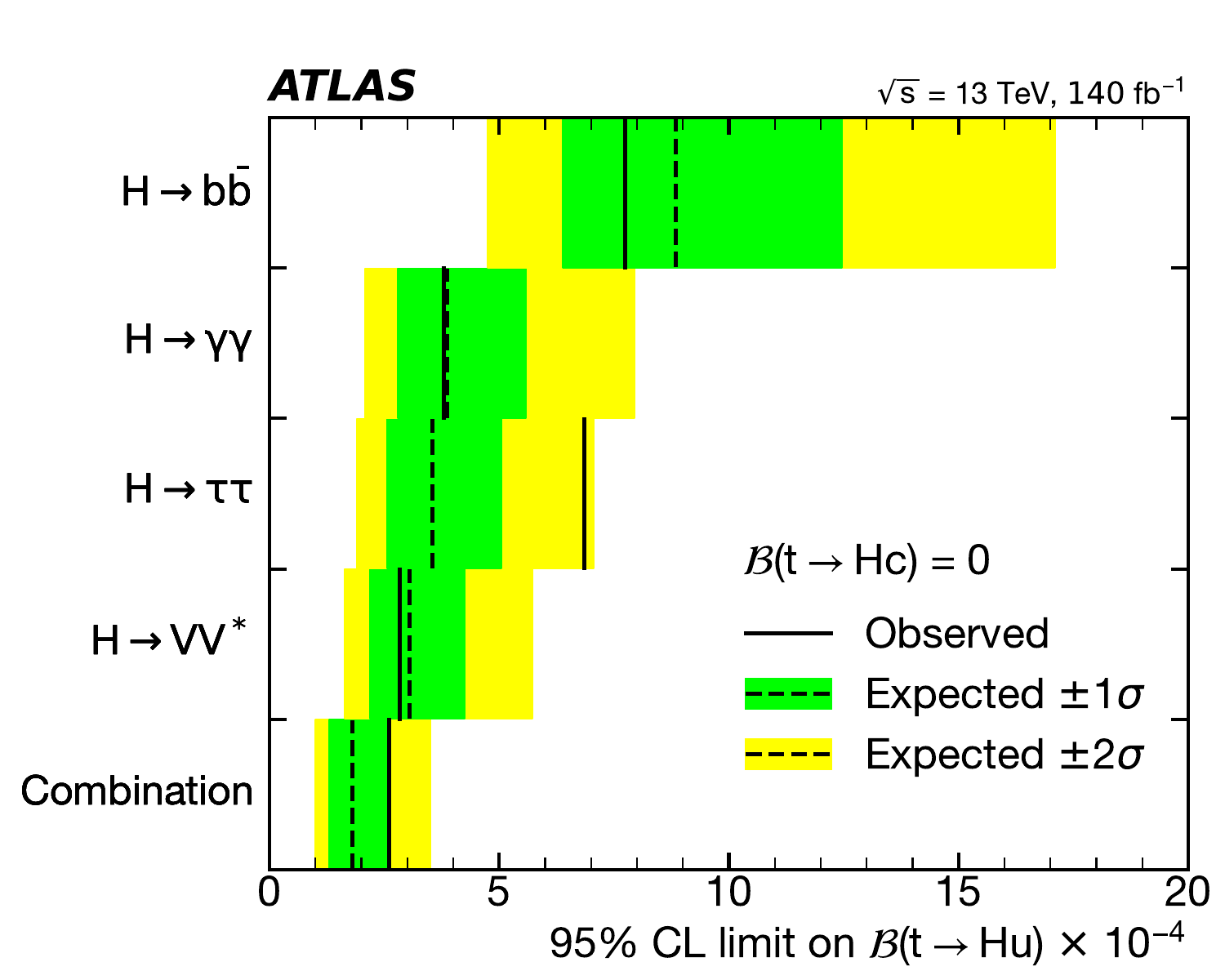}
      \caption{}
  \end{subfigure}
  \hfill
  \begin{subfigure}{0.49\textwidth}
      \includegraphics[width=\textwidth]{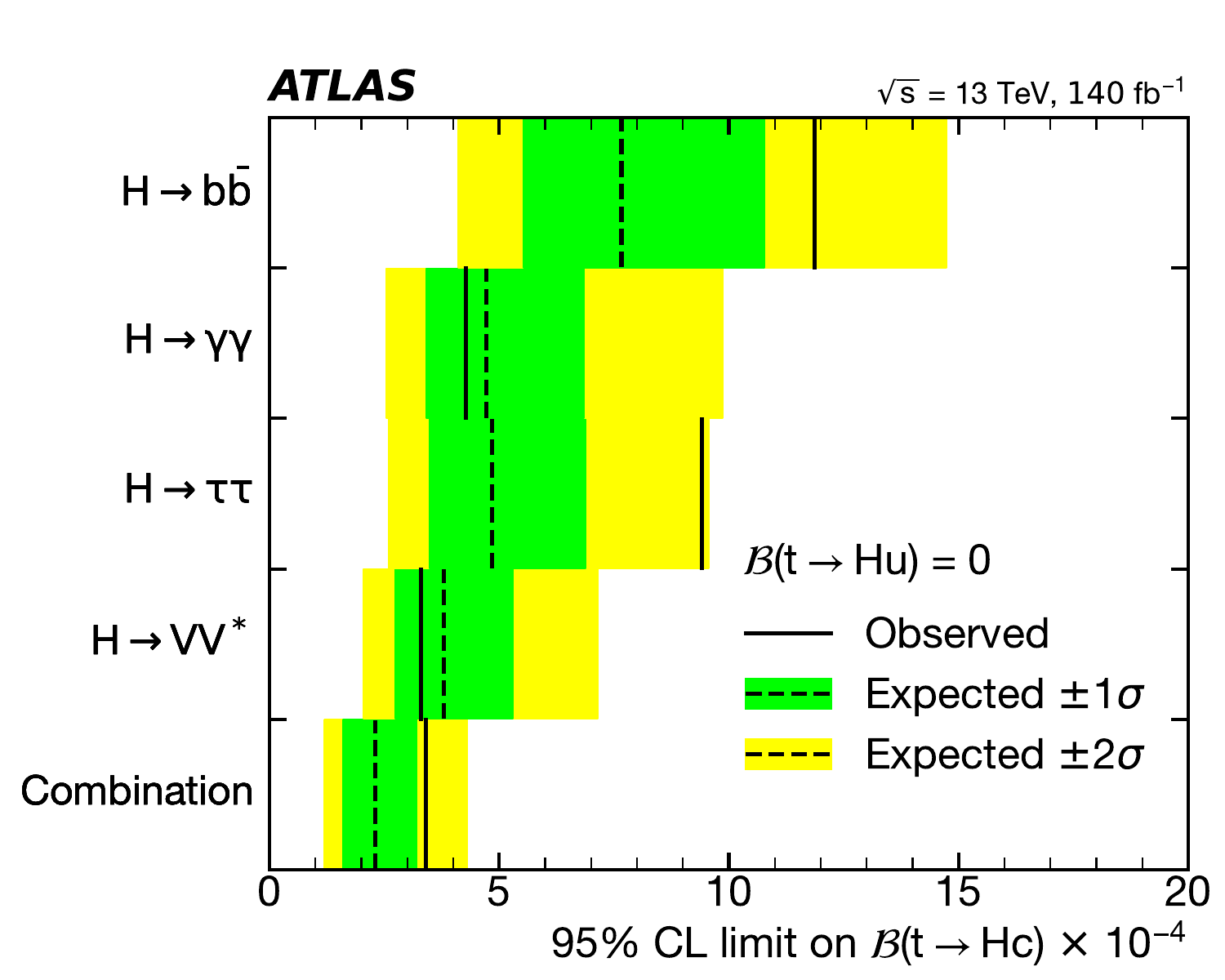}
      \caption{}
  \end{subfigure}
  \caption{
      The \SI{95}{\percent} CL upper limits on
      (a) \(\mathcal{B}(t \to Hu)\) assuming \(\mathcal{B}(t \to Hc) = 0\) and
      (b) \(\mathcal{B}(t \to Hc)\) assuming \(\mathcal{B}(t \to Hu) = 0\)
      for the individual searches and their combination.
  }
  \label{fig:combLim}
\end{figure}

These results from this search are combined with other ATLAS searches involving Higgs decay to $\tau\tau$, $b\bar{b}$, and $\gamma \gamma$. The combination improves the upper limits on branching ratios \(\mathcal{B}(t\to Hu) < 2.6 \times 10^{-4}\) and \(\mathcal{B}(t\to Hc) < 3.4 \times 10^{-4}\). The combined results are shown in~\Cref{fig:combLim}.

\section{Conclusion}
These proceedings present a search for flavour-changing neutral currents (FCNC) in the interactions between the top quark, Higgs boson, and an up-type quark, using \SI{140}{\mbox{fb\(^{-1}\)}} of LHC Run 2 data collected by the ATLAS detector. Stringent upper limits are set on the branching ratios \(\mathcal{B}(t\to Hu)\) and \(\mathcal{B}(t\to Hc)\), with no evidence of FCNC observed. These results represent the strongest constraints to date on these couplings, translating into limits on the effective field theory Wilson coefficients that describe the FCNC interactions. By combining this analysis with other ATLAS FCNC searches, further improvements in sensitivity are achieved, setting new benchmarks for future studies.

\begin{appendix}
\numberwithin{equation}{section}

\end{appendix}

\bibliography{Shayma_Wahdan_TOP2024_Proceeding}

\begin{thebibliography}{1}
\providecommand{\url}[1]{\texttt{#1}}
\providecommand{\urlprefix}{URL }
\expandafter\ifx\csname urlstyle\endcsname\relax
  \providecommand{\doi}[1]{doi:\discretionary{}{}{}#1}\else
  \providecommand{\doi}{doi:\discretionary{}{}{}\begingroup
  \urlstyle{rm}\Url}\fi
\providecommand{\eprint}[2][]{\url{#2}}

\bibitem{GIM_mechanism}
S.~L. Glashow, J.~Iliopoulos and L.~Maiani,
\newblock \emph{Weak interactions with lepton-hadron symmetry},
\newblock Phys. Rev. D \textbf{2}, 1285 (1970),
\newblock \doi{10.1103/PhysRevD.2.1285}.

\bibitem{Branco:2hdm2012}
G.~C. Branco, P.~M. Ferreira, L.~Lavoura, M.~N. Rebelo, M.~Sher and J.~P.
  Silva,
\newblock \emph{{Theory and phenomenology of two-Higgs-doublet models}},
\newblock Phys. Rep. \textbf{516}, 1 (2012),
\newblock \doi{10.1016/j.physrep.2012.02.002},
\newblock \eprint{1106.0034}.

\bibitem{ATLAS:2024mih}
{ATLAS Collaboration},
\newblock \emph{{Search for flavour-changing neutral-current couplings between
  the top quark and the Higgs boson in multi-lepton final states in 13~TeV pp
  collisions with the ATLAS detector}},
\newblock Eur. Phys. J. C \textbf{84}, 757 (2024),
\newblock \doi{10.1140/epjc/s10052-024-12994-1},
\newblock \eprint{2404.02123}.

\bibitem{PERF-2007-01}
{ATLAS Collaboration},
\newblock \emph{{The ATLAS Experiment at the CERN Large Hadron Collider}},
\newblock JINST \textbf{3}, S08003 (2008),
\newblock \doi{10.1088/1748-0221/3/08/S08003}.

\bibitem{Read:2002hq}
A.~L. Read,
\newblock \emph{{Presentation of search results: the \(CL_S\) technique}},
\newblock J. Phys. G \textbf{28}, 2693 (2002),
\newblock \doi{10.1088/0954-3899/28/10/313}.

\end{thebibliography}

\end{document}